\documentclass[twocolumn, tighten, times]{aastex63}
\usepackage{amsmath,longtable}
\usepackage[caption=false]{subfig}
\graphicspath{{./}{Figures/}}
\usepackage{comment}

\shorttitle{Disk Emission from IRAS 15398-3359}
\shortauthors{Salyk et al.}

\begin{document}
\title{CORINOS. II. JWST-MIRI detection of warm molecular gas from an embedded, disk-bearing protostar}

\correspondingauthor{Colette Salyk}
\email{cosalyk@vassar.edu}

\author[0000-0003-3682-6632]{Colette Salyk}
\affiliation{Department of Physics and Astronomy, Vassar College,
124 Raymond Avenue, Poughkeepsie, NY 12604, USA}

\author[0000-0001-8227-2816]{Yao-Lun Yang}
\affil{RIKEN Cluster for Pioneering Research, Wako-shi, Saitama, 351-0198, Japan}

\author[0000-0001-7552-1562]{Klaus M. Pontoppidan}
\affiliation{Jet Propulsion Laboratory, California Institute of Technology, 4800 Oak Grove Drive, Pasadena, CA 91109, USA}
\affiliation{Division of Geological and Planetary Sciences, California Institute of Technology, MC 150-21, 1200 E California Boulevard, Pasadena, CA 91125, USA}

\author[0000-0002-8716-0482]{Jennifer B. Bergner}
\affil{UC Berkeley Department of Chemistry, Berkeley, CA 94720, USA}

\author[0000-0003-3655-5270]{Yuki Okoda}
\affil{RIKEN Cluster for Pioneering Research, Wako-shi, Saitama, 351-0198, Japan}

\author[0000-0000-0000-0000]{Jaeyeong Kim}
\affil{Korea Astronomy and Space Science Institute, 776 Daedeok-daero, Yuseong-gu Daejeon 34055, Republic of Korea}

\author[0000-0001-5175-1777]{Neal J. Evans II}
\affil{Department of Astronomy, The University of Texas at Austin, Austin, TX 78712, USA}

\author[0000-0003-2076-8001]{Ilsedore Cleeves}
\affil{Department of Chemistry, University of Virginia,
409 McCormick Rd, Charlottesville, VA, 22904, USA}

\author[0000-0001-7591-1907]{Ewine F. van Dishoeck}
\affil{Leiden Observatory,
Leiden University, Leiden, Netherlands }
\affil{Max Planck Institute for Extraterrestrial Physics, Garching, Germany}

\author[0000-0001-7723-8955]{Robin T. Garrod}
\affil{Department of Chemistry, University of Virginia, 409 McCormick Rd, Charlottesville, VA, 22904, USA}
\affil{Department of Astronomy, University of Virginia, 530 McCormick Rd, Charlottesville, VA, 22904, USA}

\author[0000-0003-1665-5709]{Joel D. Green}
\affil{Space Telescope Science Institute, 3700 San Martin Drive, Baltimore, MD 21218, USA}

\begin{abstract}
We present James Webb Space Telescope (JWST) Mid-InfraRed Instrument (MIRI) observations of warm CO and H$_2$O gas in emission toward the low-mass protostar IRAS 15398-3359, observed as part of the CORINOS program. The CO is detected via the rovibrational fundamental band and hot band near 5 $\mu$m, whereas the H$_2$O is detected in the rovibrational bending mode at 6--8\,$\mu$m. Rotational analysis indicates that the CO originates in a hot reservoir of $1598\pm118$\,K, while the water is much cooler at $204\pm 7\,$K.  Neither the CO nor the H$_2$O line images are significantly spatially extended, constraining the emission to within $\sim$40 au of the protostar.  The compactness and high temperature of the CO are consistent with an origin in the embedded protostellar disk, or a compact disk wind.  In contrast, the water must arise from a cooler region and requires a larger emitting area (compared to CO) to produce the observed fluxes.  The water may arise from a more extended part of the disk, or from the inner portion of the outflow cavity.   Thus, the origin of the molecular emission observed with JWST remains ambiguous.  Better constraints on the overall extinction, comparison with realistic disk models, and future kinematically-resolved observations may all help to pinpoint the true emitting reservoirs.
\end{abstract}

\section{Introduction}
\subsection{Background}
Circumstellar disks are a natural outcome of the star formation process; however there are many open questions as to when mature disks -- capable of forming planets -- first emerge.  Protostellar disks, where we define {\it disks} as rotationally supported structures, likely first form at some point during the protostellar  phase (observational Class 0/I) as circumstellar accretion disks \citep{Terebey84,Shu87,Joos12}.  They  eventually evolve to the planet-forming  (observational Class II, T Tauri) disks commonly observed around pre-main sequence stars after dissipation of the protostellar envelope. The timing and rate of their formation  and evolution are uncertain, however, and the properties of the youngest disks have historically been difficult to measure.

Initially, protostellar disks were not directly detected due to the lack of spatial resolution and confusion with the protostellar envelope and outflow \citep{Belloche02,Chiang12}.  Only recently have high-resolution mm-wave interferometers been able to distinguish the Keplerian rotational signatures of disks from envelopes in the youngest targets \citep[e.g.][]{Tobin12, Murillo13,  Lindberg14, Ohashi14, Yen17, Maret20}.  The combined spatial resolution and sensitivity provided by ALMA is now allowing for sub-arcsecond characterization of line emission from a larger sample of embedded disks \citep{Ohashi23}. 

While there is a theoretical expectation that the youngest embedded disks are compact ($<$a few 10s of au; \citealp[e.g.,][]{Hueso05, Visser10,Machida11} and compilations in \citealp{Vaytet18,Tsukamoto23}), the observed statistics on embedded disk sizes find typical disk sizes of $\sim$40 au in the Class 0/I protostars of Orion, with a slight decrease in size as the protostars evolve \citep{Tobin20}. However, embedded disk sizes can also have significant variation, with some smaller than the 40 au average \citep{Ohashi23}.

 Because protostellar disks evolve into planet-forming disks, some of the initial conditions for planet-formation chemistry are set in this stage.  As is the case for more evolved disks, water is thought to be a key tracer of the chemical evolution of the youngest disks \citep{vandishoeck21}, but it has been difficult to detect. \cite{Harsono20} suggested that warm water vapor is depleted in Class I protoplanetary disks on 100\,au scales, based on non-detections of H$_2^{18}$O with ALMA,  perhaps due to sequestration into large icy grains. This is somewhat puzzling as abundant warm water in more evolved protoplanetary disks is found to be common in mid- to far-infrared spectra \citep{Carr08, Pontoppidan10,Salyk11a,Riviere-Marichalar12,Banzatti23a,Banzatti23b,Perotti23}, but a water-rich {\it inner} protostellar disk may be revealed with observations that better probe within $\sim$ 10 au \citep{Harsono20}.

Hence, observational evidence of the presence and structure of embedded protostellar disks is needed to test models of early disk formation as well as to understand the earliest stages of disk chemical evolution.  In this paper, we present detections and analysis of abundant, warm CO and water vapor in the nearby protostar IRAS 15398-3359, and we discuss its potential relationship with the embedded disk known to be present in this system.

\subsection{Target}
IRAS 15398-3359 (hereafter IRAS 15398) is a very low-mass \citep[$\lesssim0.1\,M_{\odot}$,][]{Okoda18,Thieme23} protostar in the Lupus I star-forming region at a distance of $154.9\pm3.4$\,pc \citep{Galli20}. Given its massive envelope \citep[$\sim$0.5--1.2\,$M_{\odot}$, ][]{Kristensen12,Jorgensen13}, it is likely that the system is very young. It is viewed at an inclination of $\sim 70\degr$ (\citealp{Oya14}; where i=0 means the disk is viewed face on), and has a Class 0 spectral energy distribution with a bolometric temperature of 44\,K \citep{Jorgensen13}. 

The source shows a ring-like structure in HCO$^{+}$ --- an indirect indication of its destruction by abundant water inside a radius of $\sim 150$\,au \citep{Jorgensen13}. This relatively large radius given the current luminosity suggests that a recent accretion burst has heated dust out to this radius, liberating water vapor from icy grains. \cite{Okoda18} detected the Keplerian-like signature of a compact embedded protostellar disk in IRAS 15398, as traced by 0\farcs2 ALMA imaging of Sulfur Monoxide (SO), and consistent with prior constraints from CO imaging \citep{Yen17}. With a higher resolution, ($\sim$0\farcs{1}) \citet{Thieme23} found a dust disk of $\sim$4 au with rotation signature detected in SO, where they derived lower limits on disk mass and radius of 0.022 $M_\odot$ and 31.2 au.  

Water has been directly detected in the IRAS 15398 system by Herschel-HIFI in the ground-state line \citep{Kristensen12}, and by ALMA via its isotopologues HDO and (tentatively) H$_2^{18}$O \citep{Bjerkeli16}. However, the water directly detected by ALMA is extended over a $\sim$3\arcsec\, (500\,au) region, and with a morphology indicating that it traces recently liberated ices from an outflow cavity. This reservoir is therefore not associated with the protostellar disk. At the largest scale, water ice along the line-of-sight to IRAS 15398 has been detected by Spitzer \citep{Boogert08}.

More recently, IRAS 15398 was observed with JWST as part of the COMs ORigin Investigated by the Next-generation Observatory in Space (CORINOS) program (program ID 2151, PI: Y.-L. Yang), with a detection of gaseous CO and H$_2$O reported in \citet{Yang22}.  In this work, we present further analysis of these gas-phase emission features.

\section{Data}
\subsection{Acquisition}
This paper is based on a 4.9--28\,$\mu$m spectrum obtained with the Mid-InfraRed Instrument \citep[MIRI,][]{Rieke15,Wright23} Medium Resolution Spectrometer (MRS) on JWST \citep{Gardner23}.    Obtained on 2022, July 20, the spectrum was first presented in \cite{Yang22}.  Data are available on MAST : \dataset[10.17909/qv17-1b93]{http://dx.doi.org/10.17909/qv17-1b93} . The observation uses all three MIRI-MRS sub-bands for contiguous coverage of the entire available spectral range. The 4-point dither pattern for extended sources ensures diffraction-limited resampling of the point-spread function, as well as efficient removal of bad pixels and cosmic ray impacts. The SHORT and LONG sub-bands use total exposure times of $\sim$1,400\,s, while the MEDIUM sub-band uses a total exposure time of $\sim$3,600\,s for better signal-to-noise in the bottom of the deep 10\,$\mu$m silicate feature. A dedicated background exposure was also obtained for efficient background subtraction in the presence of extended emission. 

\subsection{Reduction}
The spectrum is processed to level 3 using the JWST calibration pipeline \citep{Bushouse22}, version 1.12.5 and CRDS \citep{Greenfield16} context {\tt jwst\_1183.pmap}. The dedicated background observation was explicitly subtracted as part of the level 3 processing. One-dimensional spectra were extracted from each of the 12 sub-band cubes using an aperture with a scaled diameter of $4\times 1.22\lambda/D$, where $D=6.5$\,m. 

While the MRS spaxels are intrinsically undersampled at shorter wavelengths \citep{Wells15,Argyriou23}, the level 3 cube building step reconstructs roughly Nyquist-sampled data with spaxel sizes of $0\farcs13$ for MRS Channel 1.  The fringe pattern is reduced with a fringe reference and then the residual fringe is removed with the \texttt{residual\_fringe} task in the JWST pipeline.

\subsection{Extraction of line fluxes and correction for extinction }
\label{sec:extraction}
The continuum was subtracted from the spectrum using a robust, iterative method to estimate the continuum, as follows: The spectrum is smoothed using a median filter with a width of 15 wavelength channels. A new spectrum is defined from the previous iteration by using values that are less than those of the smoothed version of the spectrum and interpolating between them. This becomes the input spectrum to the next iteration. After three iterations, the resulting continuum is further smoothed using a second-order Savitzky-Golay filter with a box size of 15 \citep{Savitzky64}. This procedure results in a very clean continuum, under the simple assumption that all lines are in emission, as confirmed by visual inspection.  

\begin{figure*}[ht!]
\epsscale{1.2}
\plotone{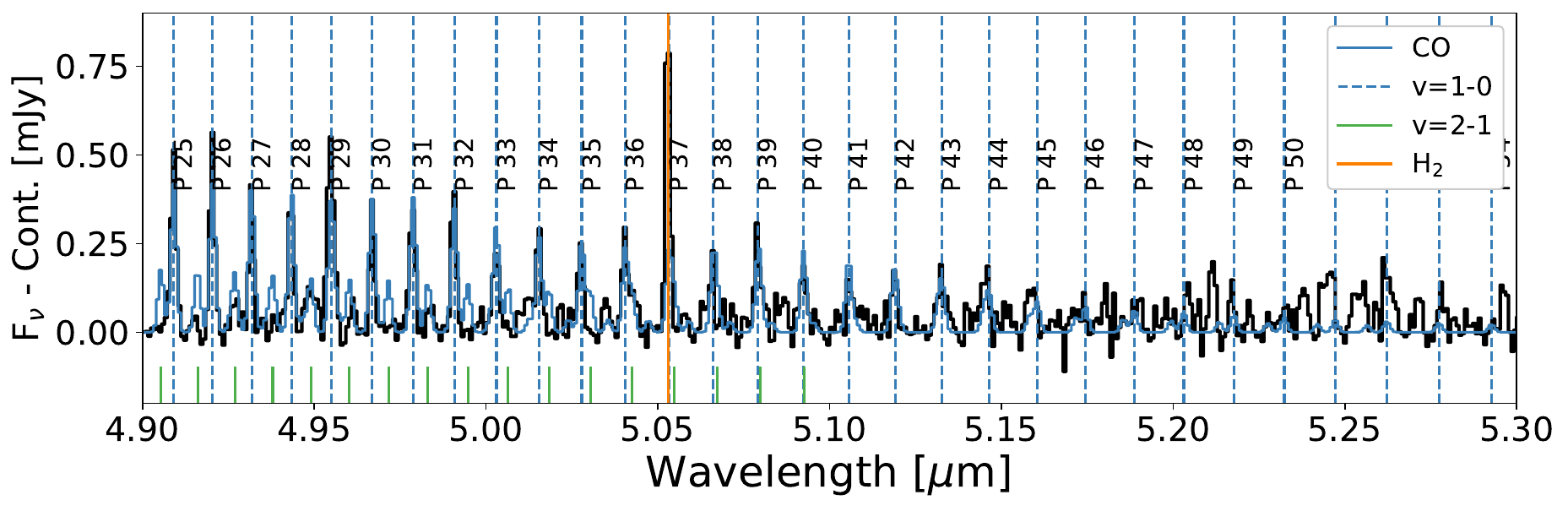}
\caption{Portion of continuum-subtracted MIRI-MRS spectrum of IRAS 15398 overlaid with an extinction-corrected $^{12}$CO emission model (blue), discussed further in Section \ref{sec:co_analysis}.  Blue vertical dashed lines and labels show the location and assignment of the $^{12}$CO v=1--0 lines; green vertical lines at the bottom mark $^{12}$CO v=2--1 lines. An orange vertical line marks the location of H$_2$ 0-0 S(8).
\label{fig:co_assignment}}
\end{figure*}

\begin{figure*}[h!]
\epsscale{1.}
\plotone{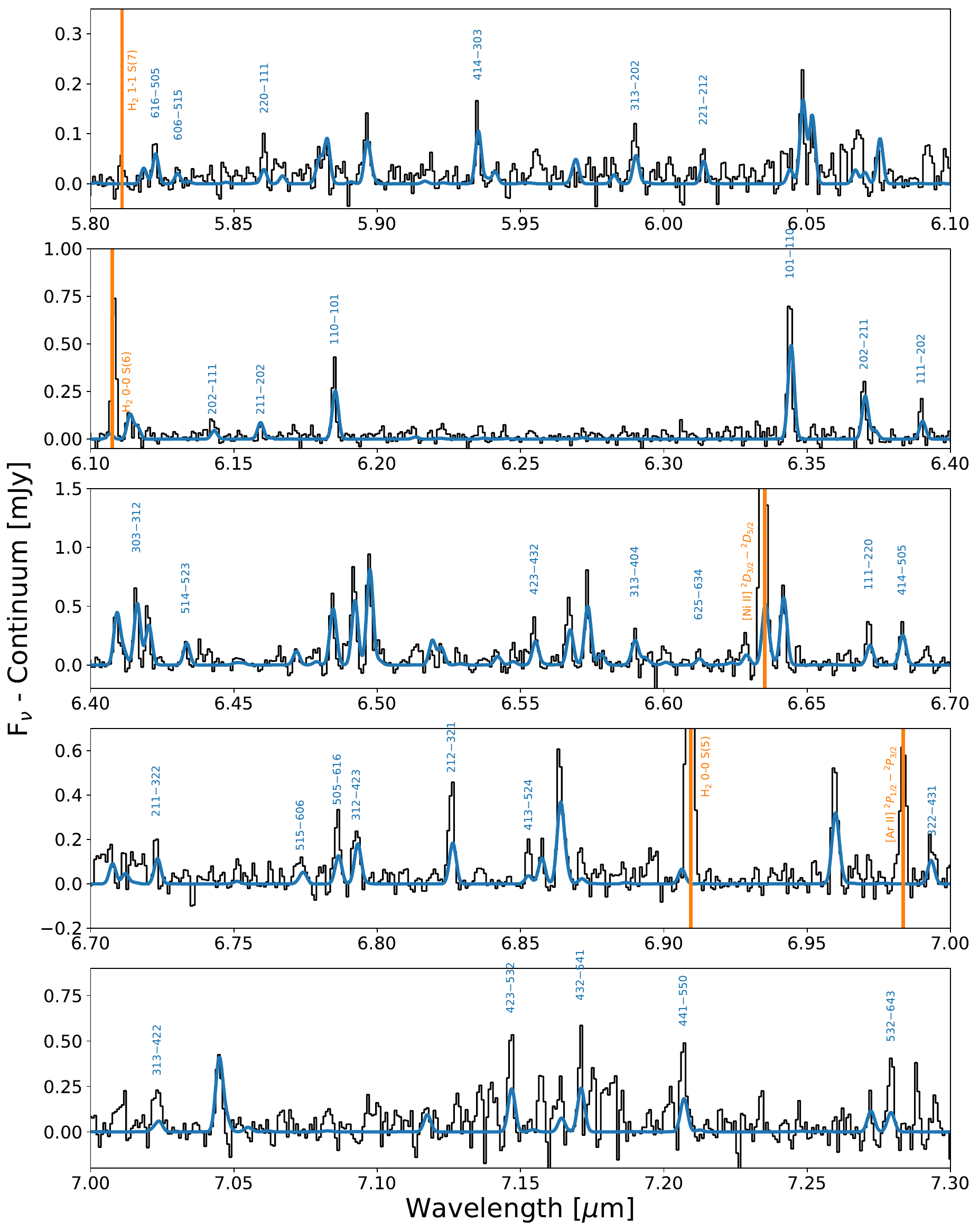}
\caption{A portion of the observed continuum-subtracted MIRI-MRS spectrum (black) compared to a maximum-likelihood water vapor ``slab'' model (blue; see Section \ref{sec:water_analysis}).  The model has been corrected for extinction and convolved to a resolution of 120 km s$^{-1}$.  Blue labels show rotational quantum numbers (J$_\mathrm{K_a K_c}$) from the HITRAN database \citep{Gordon22} for some non-blended water emission lines.  Orange lines and labels mark atomic transitions identified in \citet{Yang22}.
\label{fig:water_assignment}}
\end{figure*}

The continuum-subtracted spectrum shows emission from many CO, water, H$_2$, and atomic lines. Figure \ref{fig:co_assignment} shows the spectral region near the CO ro-vibrational fundamental (v=1--0) band.  This region shows evidence for $^{12}$CO v=1--0 and v=2--1 emission; no $^{13}$CO is detected.  Figure \ref{fig:water_assignment} shows evidence for emission from the H$_2$O ro-vibrational bending mode.  

Line fluxes are extracted using the {\it flux calculator} routine in the {\it spectools-ir} python package.  This package is available on pypi (https://pypi.org/project/spectools-ir/) and version 1.0.0 is archived on Zenodo \citep{Salyk22}.  Gaussian curves are fit to the spectra at line locations provided by the HITRAN database \citep{Gordon22}, and fits are manually vetted by the user.

Observed emission lines are heavily extinguished by a large column of dust and ice, as evidenced by the red spectral energy distribution and the presence of deep ice and silicate bands \citep[see][]{Yang22}. Thus, the line fluxes must be corrected, both in absolute and relative terms, prior to analysis. Ice extinction is particularly variable (with respect to wavelength) in the spectral region of the water vapor bending mode.  We employ a semi-empirical method to approximately correct for absolute and relative extinction. 

To correct for extinction, we assume the line emission arises from a line-emitting source surrounded by a cooler envelope.  We separate the optical depth into ice and dust components, i.e., $\tau_\mathrm{total}=\tau_\mathrm{dust}+\tau_\mathrm{ice}$.  Ice extinction is determined directly from the data, as described in \citet{Yang22}; we use a slightly updated version of the ice extinction, which fits a fourth-order polynomial baseline and considers two components of silicate dust (olivine and pyroxene) smoothed with a  Savitzky-Golay filter to isolate the ice optical depth spectrum. 

Dust extinction is estimated by modeling the H$_2$ lines together with the dust optical depth similar to the method presented in \cite{Narang23}.  The details of this method are explained in Okoda et al. (2024, in preparation).  Here we briefly introduce the approach and the modeling results.  

In rotational diagram analysis, we have that the integrated intensity from rotational state $J$, $I_J$, assuming optically thin emission, is given by
\begin{align}
    I_J = & \int I_{J,\lambda}\text{d}\lambda = N_J \frac{h c A_J}{4 \pi \lambda_J}
    \label{eq:1}
\end{align}
where $N_J$ is the column density of molecules in state $J$, $\lambda_J$ is the rest wavelength, and $A_J$ is the Einstein-A coefficient.  The column density ($N_J$) can be related to the total number of molecules ($\mathcal{N}_J$) as
\begin{align}
    N_J = & \frac{\mathcal{N}_J}{\text{emitting area}} = \frac{\mathcal{N}_J}{\Omega d^2},
\end{align}
where $\Omega$ is the solid angle of the emitting area and $d$ is the distance to the source.
Equation\,\ref{eq:1} can be rearranged and divided by the degeneracy of the state, $g_J$, to find:
\begin{align}
    \frac{N_J}{g_J} = & \frac{4 \pi \lambda_J I_J}{hcA_J g_J} \label{eq:NJgJ}
\end{align}
where $g_J$ in this case is given by $(2J+1)(2I+1)$. The total nuclear spin quantum number ($I$) is equal to 0 for even $J$ and 1 for odd $J$.  We assume that the ortho- (even) to para-H$_2$ (odd) ratio has no deviation from the statistical ratio of 3. 

$\frac{N_J}{g_J}$ is also related to the total column density of molecules, $N_\text{tot}$,
\begin{align}
    \frac{N_J}{g_J} = & \frac{N_\text{tot}}{Q(T_\text{rot})}e^{-E_{u, J}/kT_\text{rot}},
\end{align}
where $E_{u, J}$ is the upper level energy, $T_\text{rot}$ is the rotational temperature and $Q(T_\text{rot})$ is the partition function.  This can be rewritten as
\begin{align}
    ln \frac{N_J}{g_J} = & ln \frac{N_\text{tot}}{Q(T_\text{rot})} - \frac{1}{T_\text{rot}}\frac{E_{u, J}}{k}
\end{align}
by taking the natural log of both sides.  Thus, plotting $N_J/g_J$ vs. $E_{u,J}/k$ produces a line with slope $-1/T_\text{rot}$, and an intercept that depends on $N_\text{tot}$. The partition function is taken from \citet{Herbst96} as \footnote{ Use of a functional form allows for the potential separation of vibrational and rotational partition functions.  This partition function agrees with the HITRAN values to within 5\% between 500 K and 2000 K.}
\begin{equation}
    Q(T_\text{rot}) = 0.0247\, T_\text{rot}/(1-\text{exp}(-6000\,\text{K}/T_\text{rot})),
\end{equation} 
 and transition data are from \citet{Jennings84}.

\begin{figure}[htbp!]
    \centering
\includegraphics[width=0.47\textwidth]{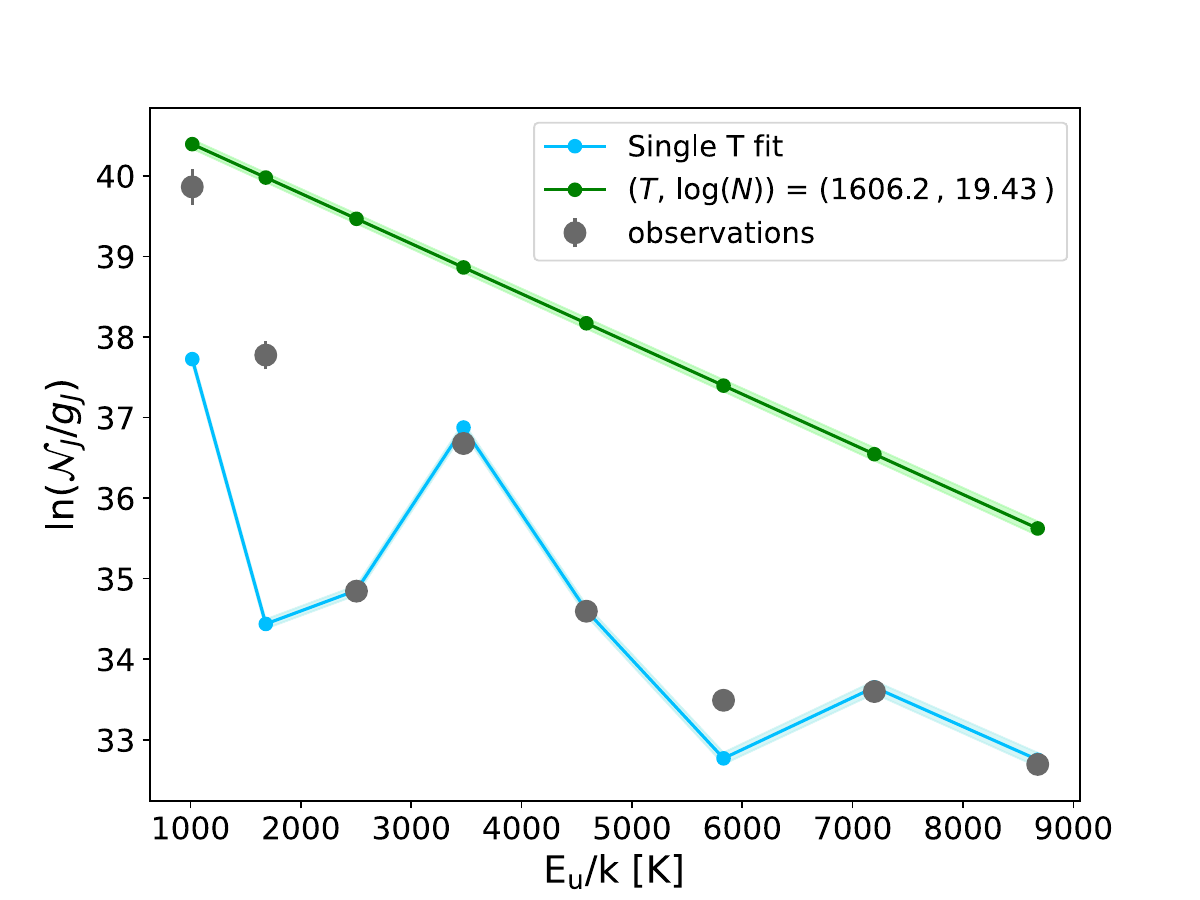}
    \caption{The H$_2$ rotational diagram toward the protostar along with the best-fitting H$_2$ model.  The H$_2$ fluxes are extracted from a single pixel (0.13\arcsec), while we measure the dust column density over a 0.77\arcsec\ aperture.  The green line shows the synthetic H$_2$ rotational diagram without dust extinction, while the blue line shows the same model with the fitted dust extinction.  The best-fitting H$_2$ excitation temperature and H$_2$ column density are 1606.2 K and 2.69$\times10^{19}$ cm$^{-2}$, respectively.
    \label{fig:H2_rot}}
\end{figure}

The IRAS 15398 spectrum covers eight H$_2$ pure rotational lines from S(1) to S(8).  Figure \ref{fig:H2_rot} shows the H$_2$ rotational diagram toward the protostar; the lower points are derived from raw fluxes and don't yet reflect the extinction correction.  The irregular shape indicates the effect of dust and ice extinction.  To model the H$_2$ lines, we can write the observed integrated H$_2$ line intensity from the $J$th level as

\begin{equation}
    I_{J, \text{obs}} = I_J e^{-(\tau_\text{ice} + \tau_\text{dust})},
\end{equation}
where $I_J$ is the intrinsic line intensity, assuming that the ice and dust are purely absorbing.  

\begin{figure*}[ht!]
\epsscale{1.}
\plotone{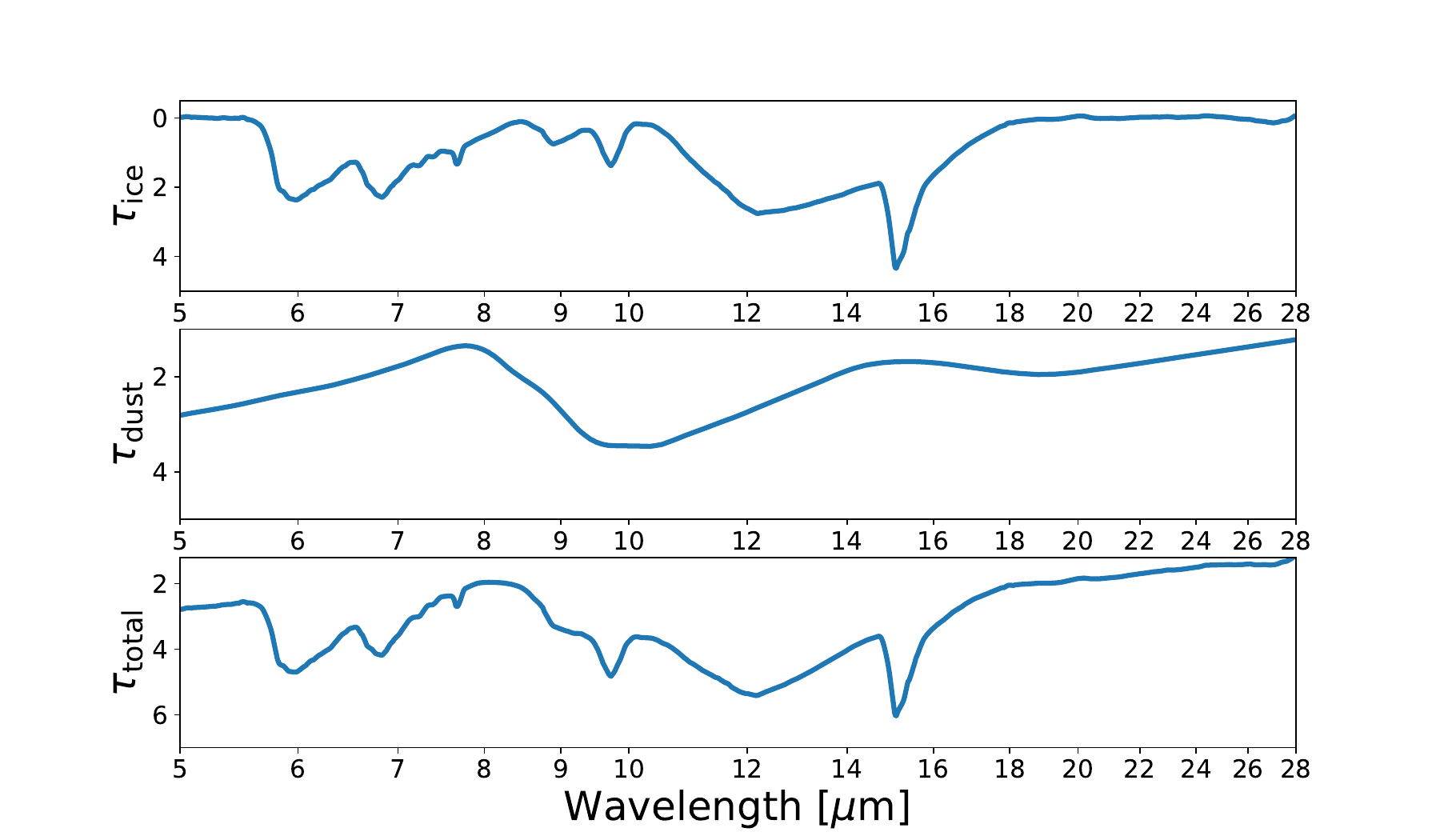}
\caption{Ice (top), dust (middle) and total (bottom) optical depth used to extinction-correct emission line fluxes.
\label{fig:extinction}}
\end{figure*}

We can construct the rotational diagram using the observed intensities and modeling $N_{J, \text{obs}}$ as $N_J\,\text{exp}(-(\tau_\text{ice} + \tau_\text{dust}))$, then solve for $N_\text{tot}$, $T_\text{rot}$ and $\tau_\text{dust}$. The $\tau_\text{dust}$ is parameterized as $\kappa_\lambda \Sigma_\text{dust}$ where $\kappa_\lambda$ is the dust opacity.  Since the ice extinction is accounted for separately, the dust-only extinction law is determined using opacities from a bare dust grain model (Bergner et al. 2024, in preparation), produced using {\it optool} \citep[][Figure\ \ref{fig:extinction}, middle]{Dominik21}.  Therefore, we can model $N_{J, \text{obs}}$ with three free parameters, $T_\text{rot}$, $N_\text{tot}$, and $\Sigma_\text{dust}$, assuming only one excitation temperature component. 

Figure \ref{fig:H2_rot} shows the best-fitting model of $N_{J, \text{obs}}(T_\text{rot}, N_\text{tot}, \Sigma_\text{dust})$ toward a single pixel closest to the protostar.  The fitting is performed throughout the MRS field of view and is discussed further in Okoda et al. (2024, in preparation).   Single-temperature fitting typically underestimates the S(1) and S(2) lines, which are better fitted with a two-temperature model.  In this analysis, we focus on deriving the dust extinction, which varies by less than 1\%\ between the single- and two-temperature models.  Also, the S(1) and S(2) lines may have contamination from diffuse cold H$_2$ gas.  Thus, we adopt here the dust column density modeled with a single temperature.  A dust column density of 1.38$\times10^{-3}$ g cm$^{-2}$ is measured from a 0.77\arcsec\ aperture, corresponding to $4\times1.22\lambda/D$ at 5 \micron, centered on the protostar  -- the same as the aperture size used in our spectral extraction. 

The fitted dust column density can be converted to an equivalent foreground envelope H$_2$ column density of 2.95$\times10^{22}$ cm$^{-2}$, assuming a mean molecular weight of 2.809 \citep{Evans22} and a 100:1 gas:dust mass ratio.  It is also equivalent to an A$_V$ of  15 mag using our dust model (A$_V$ [mag] =  5.1$\times$10$^{-22}$ N(H$_2$)).  The foreground H$_2$ column density is a factor of $\sim$ 3 lower than that derived from Herschel observations (7.2$\times10^{22}$ cm$^{-2}$; \citealp{Palmeirim13}),  somewhat consistent with the Herschel observations viewing the whole envelope, while our JWST observations see only the front.  However, it should be noted that the two studies do not use the same dust model.  

The final derived dust, ice, and total optical depth are shown in Figure \ref{fig:extinction}.  Extinction-corrected line fluxes (hereafter ``intrinsic'' line fluxes) for H$_2$O and CO are calculated using the equation 
$F_\mathrm{int}=F_\mathrm{obs}/e^{-\tau_\mathrm{total}}$,  where $F_\mathrm{int}$ is the intrinsic flux, and $F_\mathrm{obs}$ is the observed flux.  Note that optical depths in the $\sim$5--7 $\mu$m range are $\sim$ 3--5, so the resulting enhancement in intrinsic line flux as compared to observed line flux is of order $10^{2}$. Observed and intrinsic line fluxes are provided in Tables \ref{table:co_fluxes} and \ref{table:h2o_fluxes}.

\subsection{Line images}
\begin{figure*}[ht!]
    \centering
    \includegraphics[width=18cm]{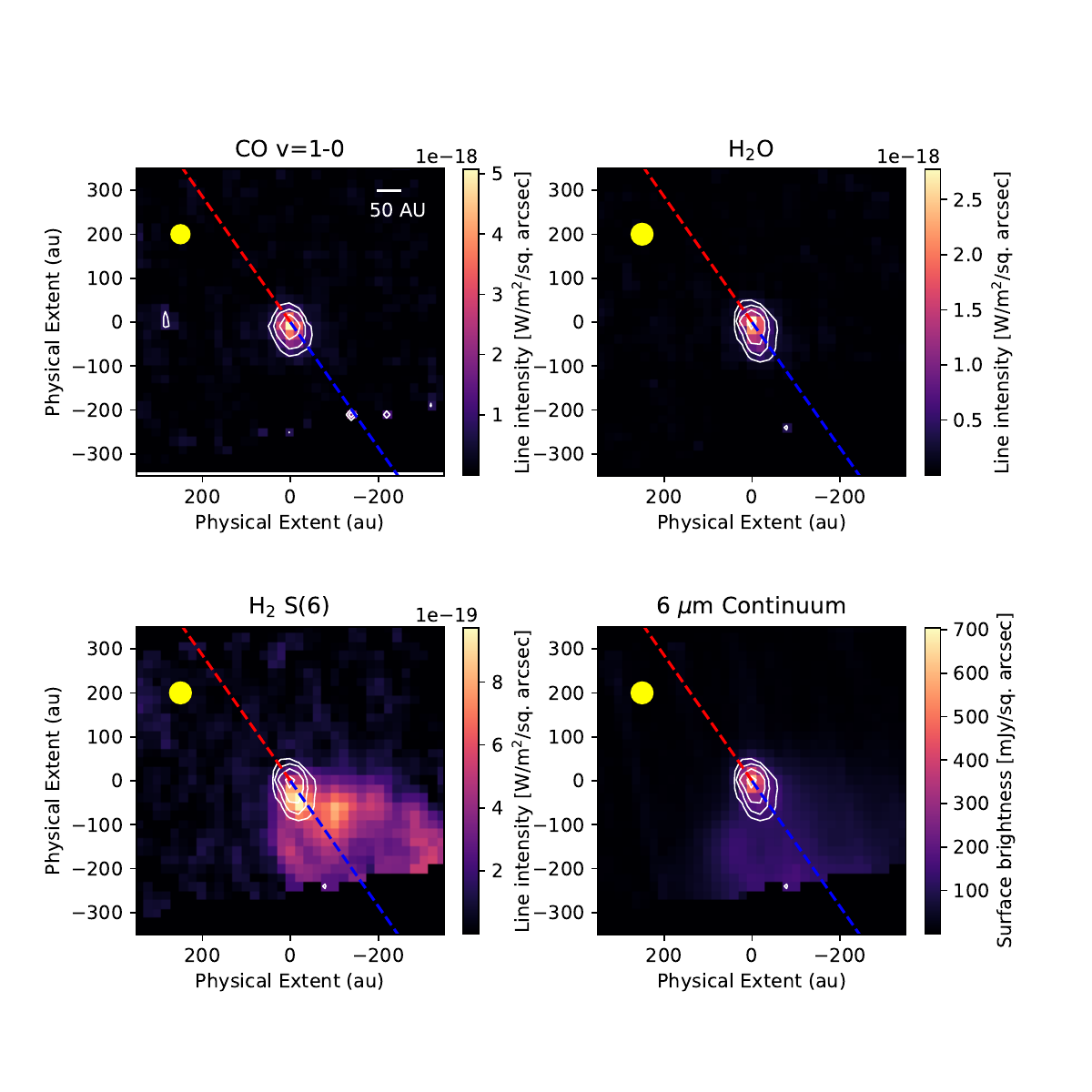}
    \caption{Images of an average of 6 CO rovibrational lines, 5 H$_2$O bending mode lines, the H$_2$ S(6) line, and the 6 $\mu$m continuum (adjacent to the H$_2$ S(6) line). In the top panels, the contours match the images at 3,6,12, and 24$\sigma$. In the bottom panels, the contours are those of the water vapor emission for comparison. The physical scale assumes a distance of 154.9\,pc. 
 Yellow reference circles have a diameter equal to the instrumental PSF FWHM from \citet{Law23}.  Red and blue dashed lines mark the red and blue outflow directions assuming PA=35$^\circ$ \citep{Bjerkeli16}.}
    \label{fig:line_images}
\end{figure*}

To obtain a direct limit on the spatial extent of the molecular line emission, continuum-subtracted images were extracted for the CO $v=1-0$ lines and the water bending mode lines. We selected six of the strongest CO lines without other line contamination (P(26)-P(28) and P(30)-P(32)), as well as five bright, isolated water lines near 6.5 $\mu$m. The lines were integrated over three spectral planes, assuming spectrally unresolved lines, and the continuum was estimated from an average of six adjacent planes, three on each side of the line. The line images are presented in Figure \ref{fig:line_images}.

The CO and water lines emit from a compact region, distinct from the large-scale blue outflow lobe traced by the H$_2$ S(6) line. The peak of the CO and water line emission appears nearly unresolved (FWHM$\sim 0\farcs3\times 0\farcs45$), compared to the JWST PSF at 6\,$\mu$m ($0\farcs3$; \citealt{Law23}). As seen in the line images, the CO and water emission is centered on the central continuum source, while the H$_2$ is offset by $\sim 20\,$au and is likely tracing the blue side of the outflow. A slight elongation may be present along the outflow axis for both CO and water; however, this is also seen in the continuum image, suggesting that it is due to scattering of the central source on large grains in the outflow cavity. It is similarly possible that the elongation is due to scattering off the exposed surface layer of a barely resolved edge-on disk, similar to that of L1527 IRS in Taurus \citep{Tobin12} or CRBR 2422 in Ophiuchus \citep{Pontoppidan04}. In summary, the line images demonstrate that the CO and water emission originate from a compact region with a radius of $<$25--40\,au, and centered on the 6\,$\mu$m continuum source. 

\section{Analysis}
\subsection{Properties of CO and Water emission columns}
\subsubsection{CO}
\label{sec:co_analysis}
\begin{figure}[ht!]
\epsscale{1.}
\plotone{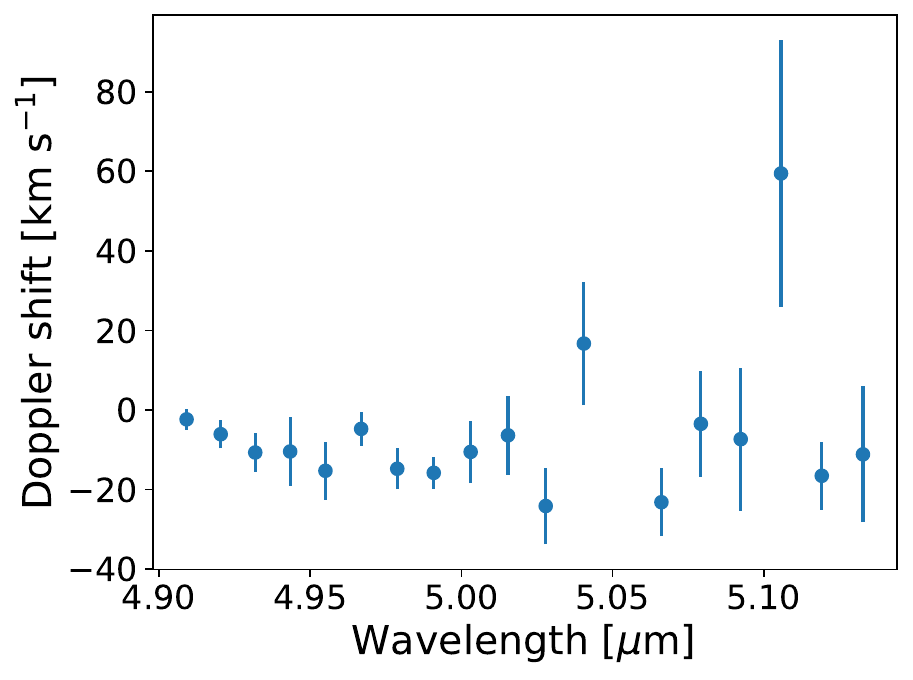}
\caption{Measured (barycentric) Doppler shifts for CO emission lines. 
\label{fig:co_doppler}}
\end{figure}

 The measured Doppler shift for CO is shown in Figure \ref{fig:co_doppler}.  The weighted mean barycentric \footnote{The default velocity reference frame for JWST data products is the solar system barycentric frame.  Barycentric Doppler shifts may differ from the heliocentric Doppler shift by up to $\sim$ 15 m s$^{-1}$ due to the sun's motion around the solar system barycenter \citep[e.g.,][]{Endl07}} Doppler shift is $-$8.3 $\pm$ 1.4 km s$^{-1}$, suggesting a $7.3$ to $7.7$ km s$^{-1}$ blueshift relative to the heliocentric source velocity of $-0.6$ to $-1$ km s$^{-1}$  (converted from $v_\mathrm{LSR}$ reported in \citealp{Jorgensen13,Bjerkeli16b}). 

\begin{figure}[ht!]
\epsscale{1.}
\plotone{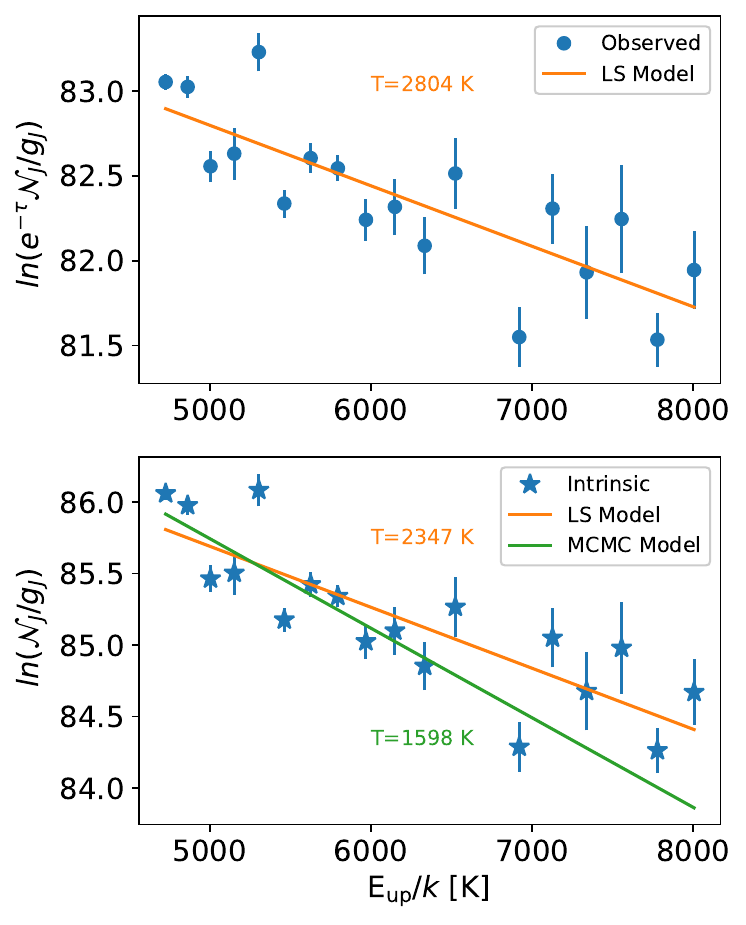}
\caption{Rotation diagram for CO emission lines.  Top: observed line fluxes, along with a least-squares (LS) model; bottom: intrinsic (extinction-corrected) fluxes with least-squares and MCMC-based models.
\label{fig:co_rotation}}
\end{figure}

A rotation diagram for CO is shown in Figure \ref{fig:co_rotation}.  Since the CO emission is only marginally spatially resolved, we work with the total line flux and a variation of Equation \ref{eq:NJgJ},
\begin{align}
     \frac{\mathcal{N}_J}{g_J} = & \frac{4 \pi d^2 \lambda_J F_J}{hcA_J g_J} 
\end{align}
where $F_J$ is the line flux and $\mathcal{N}_J$ is the total number of molecules.  As introduced in Section \ref{sec:extraction}, for optically thin emission, we can form an equation for a line, in this case,
\begin{align}
    ln \frac{\mathcal{N}_J}{g_J} = & ln \frac{\mathcal{N}_\text{tot}}{Q(T_\text{rot})} - \frac{1}{T_\text{rot}}\frac{E_{u, J}}{k}
\end{align}
such that the slope of the line is $-1/T$ and the intercept here depends on the total number of molecules, $\mathcal{N}_\text{tot}$.  Deviations from linearity can reflect temperature gradients or optical depths $>1$. 

\begin{figure}[ht!]
\epsscale{1.}
\plotone{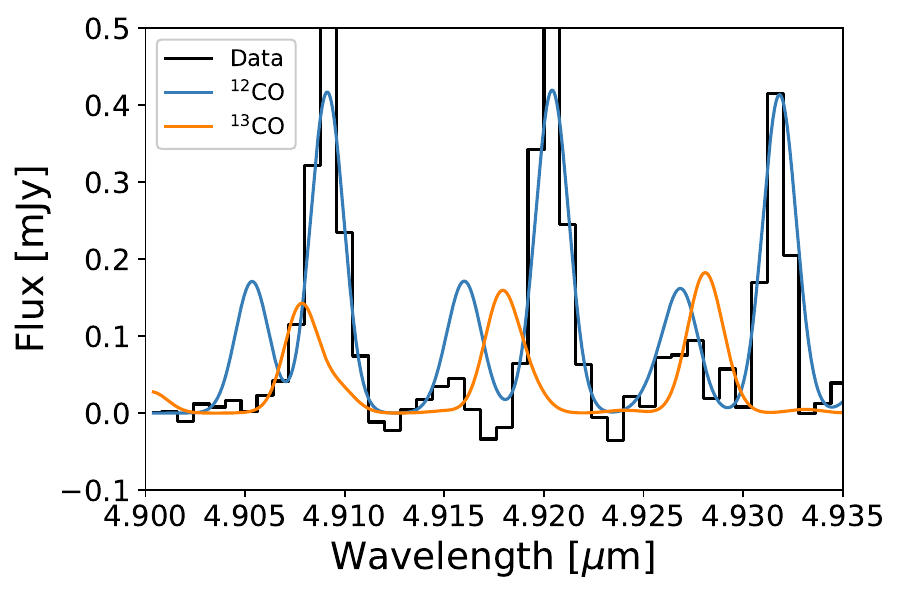}
\caption{Selected portion of the IRAS 15398 spectrum highlighting the non-detection of $^{13}$CO (orange model).  The 3:1 line strength ratio shown here is taken as our minimum detectable $^{13}$CO signature.
\label{fig:co13_nondetection}}
\end{figure}

A linear least squares (LS) fit to the observed CO rotation diagram yields a temperature of 2347 $\pm$ 380 K for extinction-corrected fluxes (versus 2804 $\pm$ 522 K for raw observed fluxes).  The linearity of the rotation diagram suggests that the emission is optically thin; however, some caution is warranted as high-J rotational lines can remain close to linear even as optical depths rise \citep{Herczeg11,Francis24}.  Therefore, we explore whether we can place any constraints on optical depth given the non-detection of $^{13}$CO.  Figure \ref{fig:co13_nondetection} shows a region of the spectrum where we would expect to observe two relatively isolated $^{13}$CO emission features.  We show a $^{13}$CO model in which the strength ratio of the 4.909 $\mu$m $^{12}$CO to the 4.918 $\mu$m $^{13}$CO lines is set to 3, and take this ratio to be a conservative estimate of the minimum observable $^{13}$CO line strength.

\begin{figure}[ht!]
\epsscale{1.}
\plotone{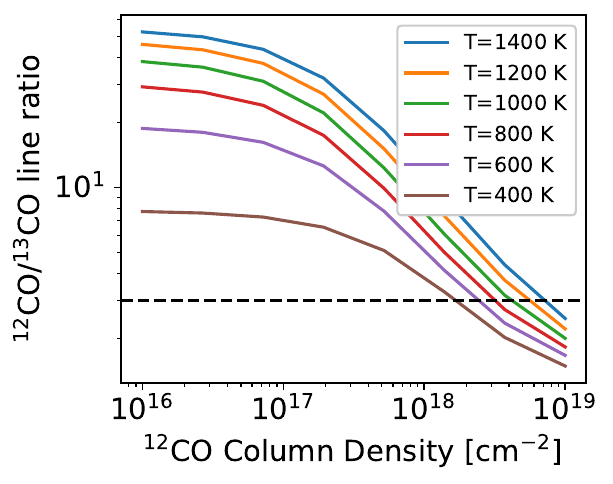}
\caption{$^{12}$CO/$^{13}$CO line ratio (see text) as a function of $^{12}$CO column density and CO model temperature.  The horizontal dashed line marks a line strength ratio of 3, our assumed minimum detectable level.  \label{fig:co13_ratio}}
\end{figure}

In Figure \ref{fig:co13_ratio}, we show how the selected $^{12}$CO/$^{13}$CO line ratio changes as we consider CO models with different column densities.  For all models, we assume that the $^{12}$CO and $^{13}$CO have the same temperature, and we assume an abundance ratio of 68 --- equivalent to that of the local ISM \citep[][and references therein]{Milam05}.  
Figure \ref{fig:co13_ratio} demonstrates that the constraint on column density provided by the $^{13}$CO non-detection has a weak temperature dependence, but that we can exclude $^{12}$CO column densities $>2\times10^{18}\,$cm$^{-2}$ for a wide range of CO temperatures.  Note that this column density is above that at which the lines begin to become optically thick; optical depths $>1$ occur when the curves begin to bend downwards ($N\sim10^{17}$cm$^{-2}$).  Therefore, both optically thin models, and some models with moderate optical depth, are consistent with the $^{13}$CO non-detection.   It should also be noted that  $^{12}$C/$^{13}$C ratios measured thus far in  Young Stellar Objects and dense clouds are higher than the ISM value, ranging from 85-167 \citep{Lambert94,Federman03,Goto03,Smith15}.  With no way to independently determine this ratio for IRAS 15398, we note that a higher $^{12}$C/$^{13}$C ratio would raise the maximum allowed CO column density in proportion to its increase over the ISM value.

Using these constraints on column density, we then fit the CO emission with a Markov Chain Monte Carlo (MCMC) sampler using the {\it slab\_fitter} routine in {\it spectools-ir} \citep{Salyk22}. This routine utilizes the sampler ``emcee'' \citep{Foreman-Mackey13} with flat priors to fit observed line fluxes with a ``slab'' gas model, which assumes that the emission arises from a slab of gas with a single temperature ($T$), column density ($N$) and solid angle ($\Omega$).  We extend our uniform priors from 300 K to 3000 K for temperature to allow for a wide range of temperatures.  For column density $N$, we allow log $N$ to extend from 14 to 18.3 (cm$^{-2}$), which accommodates the optically thin regime on the low end, and restricts higher values according to the $^{13}$CO non-detection.  For solid angle, we assume log $\Omega$ is between -18 and -11.3, which are equivalent to circular emitting radii of 0.01 and 40 au (the latter a constraint provided by our spatial images) at a distance of 154.9 pc.

\begin{figure*}[htbp!]
    \centering
\includegraphics[width=0.47\textwidth]{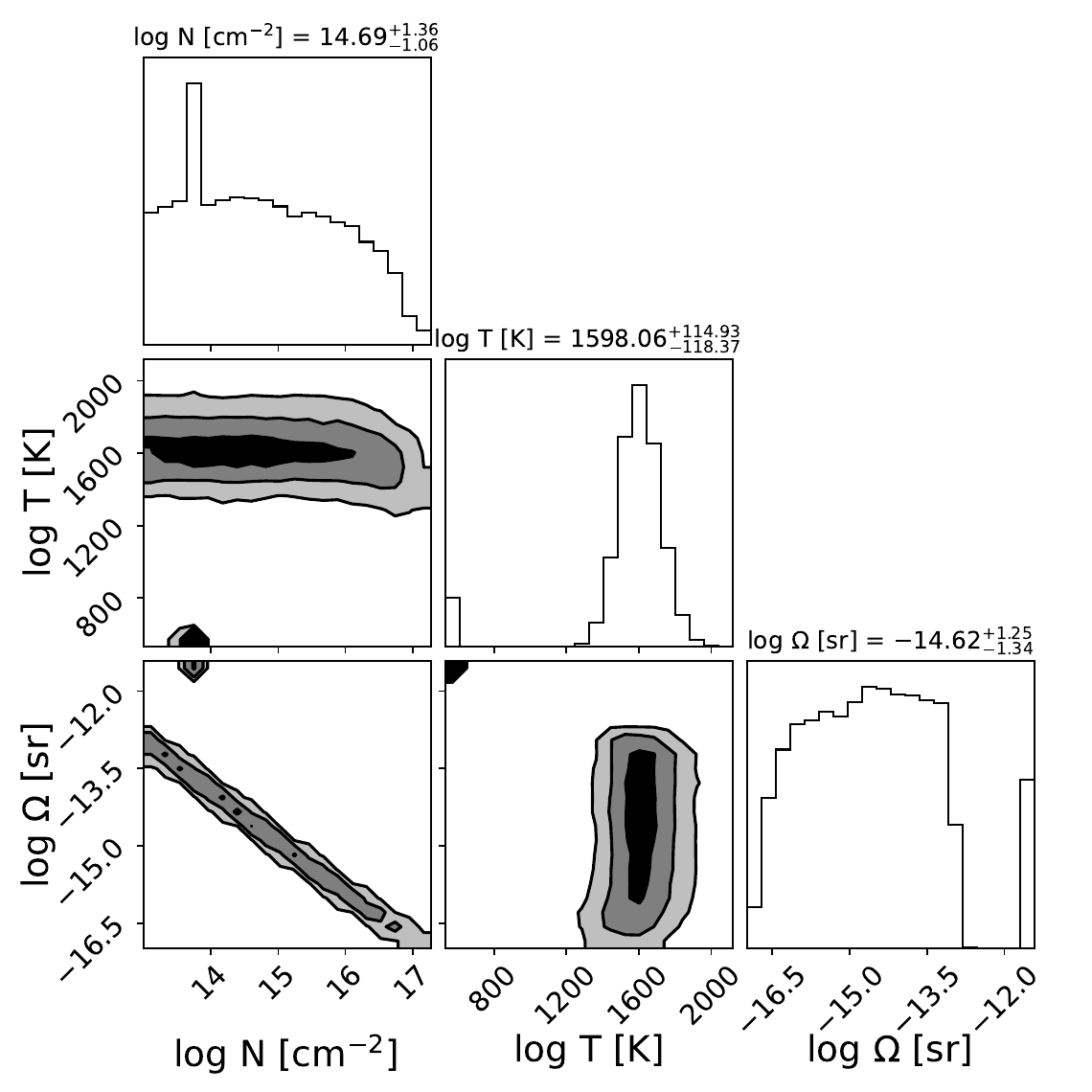}
    \caption{Corner plot for MCMC fit to $^{12}$CO emission.  Contours show 1, 2 and 3$\sigma$.}
    \label{fig:iras_corner}
\end{figure*}

A corner plot for the three free parameters is shown in Figure \ref{fig:iras_corner}.  The sampler prefers an optically thin fit, and thus the corner plot shows a degeneracy between $N$ and $\Omega$, but a constrained value of $T$.  (The sampler also spends some time in a small part of low-$T$ parameter space, but we find that these provide a poor fit to the high-J rotational line fluxes).  The best-fit temperature (which we take as the median of the posterior) is 1598 $\pm$ 118 K.  We show a best-fit model in the bottom panel of Figure \ref{fig:co_rotation}, and in Figure \ref{fig:co_assignment}.  Note that the MCMC-derived temperature is somewhat lower than what the least-squares fit to the rotation diagram provided. This likely arises due to differences in the minimization process: the least-squares fit is a fit to the y values on the rotation diagram, and so both strong and weak lines end up with similar weights; in addition, least-squares fitting minimizes distance from the line rather than $\chi^2$.  The MCMC-derived fit minimizes the line flux residuals and properly takes into account the line flux errors by minimizing $\chi^2$.

The product of $N$ and $\Omega$ also provides the total CO mass, after accounting for distance.  Using the median of the sampler output, we find a CO mass of  1.3$\times10^{19}$ g, or log $M_\mathrm{CO} = 19.1\pm0.1 $ including the 1$\sigma$ width of the sample distribution. This is equivalent to a total H$_2$ gas mass in the CO emitting layer of $ 4.6\times10^{-12} M_\odot$ assuming an H$_2$:CO abundance ratio of $10^4$ (and also accounting for the 2:28 mass ratio for the two molecules).

Figure \ref{fig:co_assignment} also demonstrates the presence of several $^{12}$CO v=2--1 emission lines. Although their low line/continuum ratio precludes a separate analysis, the model is similar to or slightly higher than the data. 
 Similarity between this thermal model and the data would be consistent with similar vibrational and rotational temperatures (i.e., the vibrationally-excited v=2 state is consistent with being thermally populated), while differences would imply non-thermal excitation of the different vibrational states.

\subsubsection{Water}
\label{sec:water_analysis}
\begin{figure}[ht!]
\epsscale{1.}
\plotone{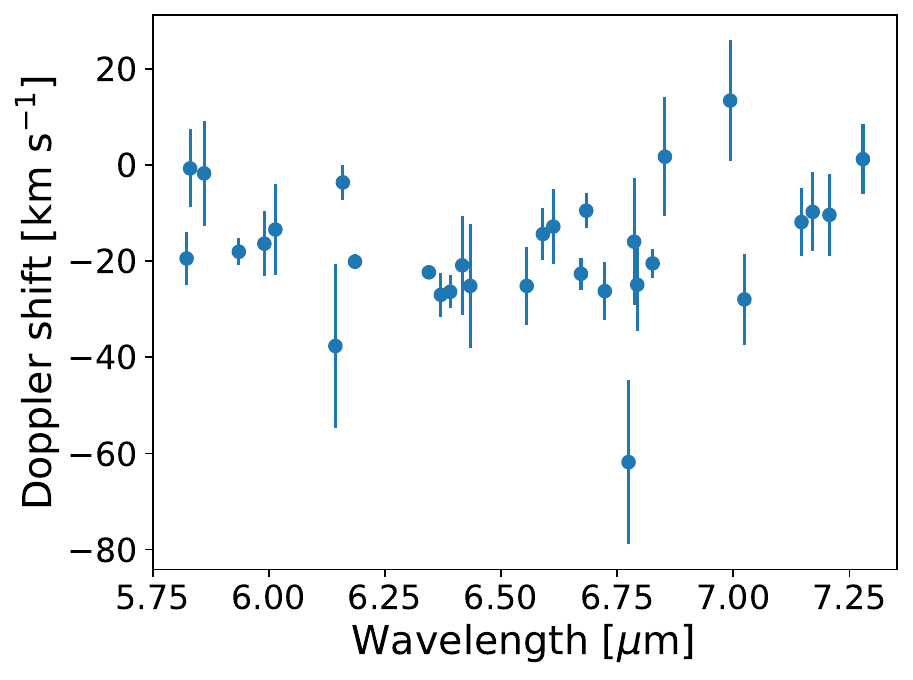}
\caption{Measured (barycentric) Doppler shifts of water emission lines.
\label{fig:water_doppler}}
\end{figure}

The measured Doppler shifts for non-blended H$_2$O lines are shown in Figure \ref{fig:water_doppler}.  The weighted mean Doppler shift is $-19.43 \pm 0.71$ km s$^{-1}$ --- blueshifted relative to the $-0.6$ to $-1$ km s$^{-1}$ heliocentric source velocity by $\sim$ 18 km s$^{-1}$.

\begin{figure}[ht!]
\epsscale{1.}
\plotone{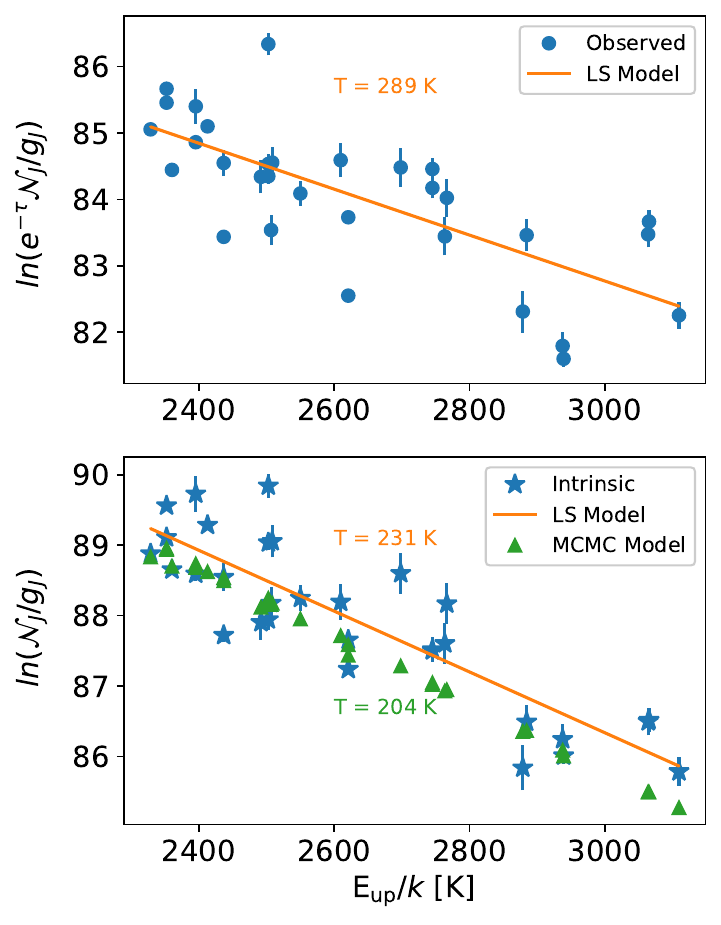}
\caption{Rotation diagram for (non-blended) water emission lines.   Top: observed line fluxes, along with a least-squares (LS) model; bottom: intrinsic (extinction-corrected) fluxes with least-squares and MCMC-derived slab models.
\label{fig:water_rotation}}
\end{figure}

A rotation diagram for H$_2$O is shown in Figure \ref{fig:water_rotation}.  In this diagram, and in our analyses, we assume an ortho/para ratio of 3,  as with this assumed value, there is no observed vertical offset between ortho and para lines in the rotation diagram, as would occur if the assumed ortho/para ratio had another value. The water emission lines lie on top of both water ice and methanol ice bands \citep{Yang22}, so extinction can substantially alter relative line fluxes. In addition, the observed fluxes show a ``raining down'' of points (i.e., vertical deviations from linearity) which can be a possible signature of high optical depth \citep{Banzatti23a}, as more optically thick lines produce less flux than in the optically thin limit.  Therefore, the raw fluxes and associated least-squares fit should not be taken as a reliable estimate of true gas properties.  

After extinction correction, the rotation diagram appears closer to a straight line, which would imply a lower column density (closer to optically thin emission).  In addition, we find that deviations from linearity are not correlated with optical depth, suggesting that the ``raining down'' of points is not due to optical depth effects but is likely random noise.  A linear least squares fit to the rotation diagram for the extinction-corrected fluxes yields a temperature of 231 $\pm$ 25 K (see Figure \ref{fig:water_rotation}). 

\begin{figure}[ht!]
\epsscale{1.}
\plotone{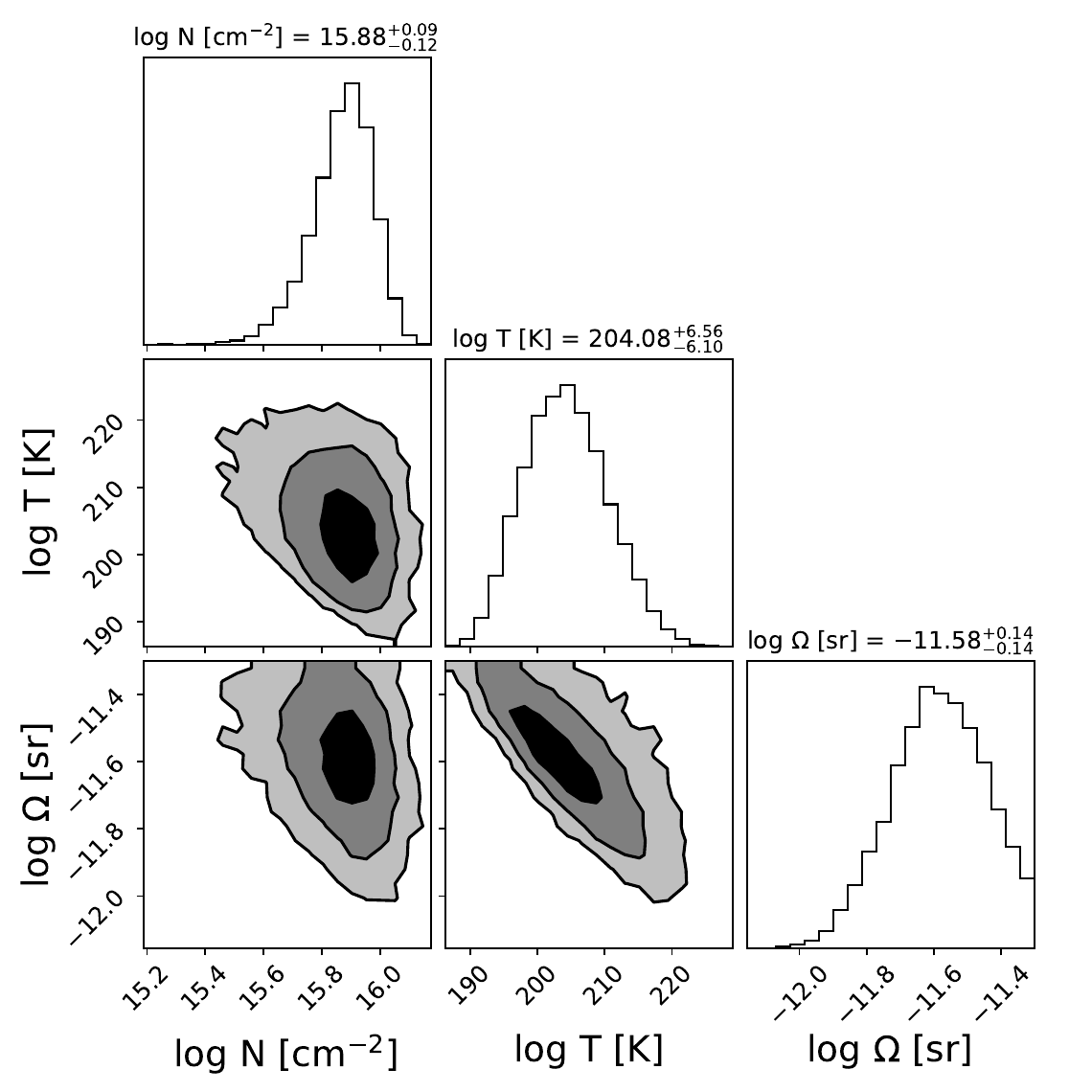}
\caption{Corner plot for MCMC analysis of water vapor emission lines.  Contours show 1, 2 and 3$\sigma$.
\label{fig:h2o_corner}}
\end{figure}

 We also fit the data using an MCMC sampler, following the same procedure as for CO.  We used flat priors with $T$ varying from 10 to 1000 K, log N varying from 9 to 18 (cm$^{-2}$) and log $\Omega$ varying from -15 to -11.3 (the latter again equivalent to a circular emitting radius R$\sim$40 au).  A corner plot is shown in Figure \ref{fig:h2o_corner}.  We find that the posterior peaks at parameters log $N =  15.88 \pm 0.12$ (equivalent to $ 7.6 \times 10^{15}$ cm$^{-2}$), $T=  204\pm 7 $ K and log $\Omega =  -11.58\pm0.14$, corresponding to a circular emitting radius of $28.9$ au.   The associated total H$_2$O mass is 1.4$\times10^{23}$ g. 

The MCMC-based model is shown on the bottom rotation diagram in Figure \ref{fig:water_rotation}, and as a spectrum in Figure \ref{fig:water_assignment}.  The model is nearly linear on the rotation diagram, though slight non-linearities arise because the strongest lines are becoming optically thick at this column density (with maximum $\tau\sim 0.9$).  The corner plots show a slight tension between the models preferred by the fluxes, and the prior constraint on $\Omega$ provided by the line images; this manifests as the posterior distribution peaking right at the upper edge of the allowed $\Omega$ values.  The sharp cutoff at large $\Omega$ (and by extension, at small $N$ and $T$ ) reflects the constraints placed on the prior. 

\section{Discussion}
In this work, we analyze newly-discovered mid-IR molecular gas-phase emission from the protostar IRAS 15398 -- a discovery only made possible with the exquisite sensitivity of JWST's MIRI-MRS instrument.  Prior to JWST, gas phase emission was commonly detected from Class I and II protostars \citep[e.g.][]{Najita03, Herczeg11}, and attributed to disk atmospheres at radii of a few au \citep[e.g.][]{Carr08,Salyk08}.  IRAS 15398 shows rotation signatures in SO, with a centrifugal barrier estimated at $\sim$ 40 au \citep{Okoda18}.  The discovery of molecular emission from IRAS 15398 therefore prompts the question --- does this emission arise from the protostar's developing disk?

The MIRI-MRS line images constrain the observed CO and H$_2$O emission to R$\lesssim$40 au, consistent with the size of the SO disk \citep{Okoda18}, and in contrast to the much larger outflow cavity traced by HDO \citep{Bjerkeli16} in ALMA images.  However, the CO and H$_2$O emission observed from IRAS 15398 do not have exactly the same characteristics as emission observed from Class II disks. 

The CO emission temperature of $>1500$ K is similar to CO observed from Class II T Tauri disks, in which the emission arises at or near the dust sublimation radius \citep{Salyk11b,Banzatti15}.  However, the emission observed from Class II disks is consistent with $N\gtrsim10^{18}$  cm$^{-2}$ \citep{Salyk11b}, at least a factor of a few higher than would be consistent with our observation of optically thin CO.  The CO mass we derive here is also a factor $10^2-10^4$ lower than total CO masses derived for the emitting column of T Tauri disks \citep{Salyk11b}.  If the CO we observe here arises from the inner circumstellar disk, perhaps this young disk atmosphere is less settled than in Class II disks, revealing a smaller CO gas column above the dust $\tau_{5 \mu \mathrm{m}}=1$ layer.   This is qualitatively consistent with first results from the ALMA eDisk program, which finds evidence that dust in protostellar disks is less vertically settled than in Class II disks \citep{Ohashi23}.  The CO emission from IRAS 15398 shows a $\sim$7 km s$^{-1}$ blueshift, similar in magnitude to CO emission line blueshifts attributed to disk winds in Class II disks \citep{Bast11,Pontoppidan11}.  Therefore, the CO emission may also be associated with a slow molecular wind originating from the disk surface.

The H$_2$O emission from IRAS 15398 has different properties from the H$_2$O emission reported for Class II disks.  We derive a water temperature of $\sim200$ K, in contrast to typical $\sim500$ K temperatures observed from Class II disk atmospheres \citep[e.g.][]{Salyk11a}.  However, the improved spectral resolution of JWST as compared to Spitzer-IRS is now revealing multiple temperature components in water emission spectra from Class II disks, with cooler components as low as $\sim$ 200 K \citep{Banzatti23b,Gasman23}. \citet{Banzatti23b} suggest that the cooler water component might arise near the water snowline.  

 For IRAS 15398, MCMC modeling reveals a preferred water emission model with a radius close to the 40 au limit provided by the image spatial extent. This is much larger than Class II disk water-emitting radii of a few au \citep[e.g.][]{Carr08,Salyk15}.  It is, however,  consistent with the water ice sublimation radius of $\sim$35 au (Bergner et al. 2024, in preparation) in the inner envelope, and the general observation that protostellar (Class 0/I) disks are hotter than their Class II counterparts \citep{vantHoff20}.  Therefore, the disagreement with typical Class II disk properties does not preclude a disk origin for the water.  However, the observed water may alternatively be associated with the ALMA-observed water, which is attributed to desorption at the inner edge of the outflow cavity \citep{Bjerkeli16}.  That emission is considerably more extended --- out to 500 au --- but perhaps the higher excitation mid-infrared lines probe a warmer portion of this region.  The observed $\sim$18 km s$^{-1}$ blueshift for the water emission may also be consistent with an outflow origin. 

 Herschel-HIFI observations of water in IRAS 15398 show broad emission profiles ($\Delta v_\mathrm{max}$ = 28 km/s) usually associated with shocks in the outflow \citep{Kristensen12}. Indeed, spatially extended far-IR water emission from IRAS 15398 with Herschel-PACS is found to be extended along the outflow \citep{Karska13}. Median water excitation temperatures based on the Herschel data are $\sim$140 K \citep{vandishoeck21} although that for IRAS 15398 is only 50 K \citep{Karska13}. In all cases, water is thought to be subthermally excited, so these excitation temperatures do not reflect the true gas temperatures \citep{Herczeg12}.  The CO excitation temperatures corresponding to the same gas are 300-700 K, consistent with shock-heated gas with kinetic temperatures up to 1500 K.

For one source, NGC 1333 IRAS4B, \citet{Watson07} had suggested that the water mid-infrared emission with T$_\mathrm{ex}\sim200$ K observed with Spitzer arises from an accretion shock onto the young disk at the disk-envelope boundary. Based on the Herschel-PACS spectrum, however, \citet{Herczeg12} suggest instead that this warm water is offset from the source and arises from outflow shocks. Indeed, new spatially-resolved JWST data find that the hot mid-infrared CO and water emission is clearly offset by $\sim$4'' (van Dishoeck et al. 2024, in preparation), demonstrating that there is no evidence for any relation with a disk accretion shock.
 
If the $\sim1500$ K CO emission arises from the inner disk, it is also curious that no warmer water component, analogous to that seen in Class II disks, is seen in IRAS 15398.  Perhaps IRAS 15398's high inclination shields the warm few au region from view; if so, we should expect the warm water component to appear in less-inclined systems.  Alternatively, perhaps the inner disk radiation environment dissociates water while leaving CO intact.  This could be caused by water's ability to dissociate over a broader UV range as compared to CO \citep{Heays17}.  Indeed, water-poor yet sometimes CO-rich infrared spectra are observed around Class II disks with inner disks depleted in small dust grains \citep{Salyk15, Perotti23}.

The slight tension between the preferred size of the emitting region according to the line fluxes, and that disallowed by the spatial extent may indicate excitation of the upper vibrational state above that expected from an LTE model, which would allow more flux to be emitted from a small area.  This could be caused by infrared pumping into the $v=1$ level, as recently observed for SO$_2$ emission \citep{vanGelder23}.  This could also explain the lack of observed water emission in the pure rotational lines.  If radiative effects are being observed, the water emission may arise in a cooler region than that being derived via an LTE assumption.

This work also highlights an important difference relative to past work on Class II disks --- the extinction must be well-characterized, as uncertainties in extinction can affect determination of gas physical parameters.  The extinction has the largest effect on determinations of molecular mass, since the mass scales (linearly, if optically thin) with the observed intrinsic flux.  The extinction correction also influences temperature, albeit more subtly, by changing the slope of line flux vs. wavelength.  Changes in temperature, in turn, influence the mass or column density required to produce the observed flux.  This problem was not encountered for the analysis of Class II disk atmospheres due to their minimal infrared extinction.  Better constraints on absolute extinction and extinction laws will be necessary to properly model molecular emission from embedded targets.

Confirmation that the observed molecular emission indeed arises from a disk would open a window into studying disk chemistry around the youngest protostars, potentially allowing for the study of disk chemical evolution through time.  However, the very different temperatures derived for the CO and H$_2$O from IRAS 15398 suggest different physical origins for the two molecules. Therefore, the observations of these two molecules cannot as yet be used to measure relative chemical abundances in this planet-forming disk.  The lack of detectable hot water does, nevertheless, suggest that the inner disk is water poor.

\acknowledgements
This work is based on observations made with the NASA/ESA/CSA James Webb Space Telescope. The data were obtained from the Mikulski Archive for Space Telescopes at the Space Telescope Science Institute, which is operated by the Association of Universities for Research in Astronomy, Inc., under NASA contract NAS 5-03127 for JWST. These observations are associated with JWST GO Cycle 1 program ID 2151. Y.-L.Y. acknowledges support from Grant-in-Aid from the Ministry of Education, Culture, Sports, Science, and Technology of Japan (20H05845, 20H05844, 22K20389), and a pioneering project in RIKEN (Evolution of Matter in the Universe). A portion of this research was carried out at the Jet Propulsion Laboratory, California Institute of Technology, under a contract with the National Aeronautics and Space Administration (80NM0018D0004).

\clearpage

\appendix

\section{Line Fluxes}
 We provide CO and H$_2$O line fluxes in Tables \ref{table:co_fluxes} and \ref{table:h2o_fluxes}.

\begin{deluxetable*}{ccccccc}[b]
\tabletypesize{\footnotesize}
\tablecolumns{7}
\tablewidth{0pt}
\tablecaption{ CO Line Fluxes \label{table:co_fluxes}}
\tablehead{
\colhead{$\lambda_0$ }& \colhead{Trans.} &
 \colhead{E$_\mathrm{up}/k$} & \colhead{Observed Flux} & \colhead{Error} & 
\colhead{Intrinsic Flux} & \colhead{ Error}\\
\colhead{[$\mu$m]}&&\colhead{[K]}&\colhead{[$\mathrm{W\,m^{-2}}$]}&\colhead{[$\mathrm{W\,m^{-2}}$]}&\colhead{[$\mathrm{W\,m^{-2}}$]}&\colhead{[$\mathrm{W\,m^{-2}}$]}}
\startdata
4.9091 & P 25 & 4725 & 1.24e-19 & 5.53e-21 & 2.51e-18 &  1.12e-19 \\
4.9204 & P 26 & 4862 & 1.24e-19 & 8.09e-21 &  2.38e-18 &  1.55e-19 \\
4.9318 & P 27 & 5003 & 8.01e-20 & 7.43e-21 &  1.47e-18 &  1.36e-19 \\
4.9434 & P 28 & 5151 & 8.86e-20 & 1.37e-20 &  1.57e-18 &  2.42e-19 \\
4.9550 & P 29 & 5304 & 1.66e-19 & 1.89e-20 & 2.87e-18 &  3.28e-19 \\
4.9668 & P 30 & 5462 & 6.95e-20 & 5.75e-21 &  1.19e-18  &  9.83e-20 \\
4.9788 & P 31 & 5625 & 9.30e-20 & 8.04e-21 & 1.56e-18  & 1.35e-19 \\
4.9908 & P 32 & 5794 & 8.95e-20 & 6.81e-21 & 1.47e-18  &  1.12e-19 \\
5.0031 & P 33 & 5968 & 6.75e-20 & 8.41e-21 &  1.09e-18  &  1.36e-19 \\
5.0154 & P 34 & 6148 & 7.44e-20 & 1.24e-20 &  1.20e-18  &  2.00e-19 \\
5.0279 & P 35 & 6333 & 6.03e-20 & 1.01e-20 &  9.58e-19  &  1.61e-19 \\
5.0405 & P 36 & 6523 & 9.40e-20 & 1.95e-20 &  1.47e-18  &  3.06e-19 \\
5.0661 & P 38 & 6920 & 3.71e-20 & 6.50e-21 &  5.74e-19  &  1.01e-19 \\
5.0792 & P 39 & 7126 & 8.04e-20 & 1.67e-20 &  1.25e-18  &  2.59e-19 \\
5.0924 & P 40 & 7338 & 5.61e-20 & 1.54e-20 &  8.74e-19  &  2.39e-19 \\
5.1057 & P 41 & 7555 & 7.80e-20 & 2.49e-20 &  1.20e-18  &  3.83e-19 \\
5.1191 & P 42 & 7777 & 3.88e-20 & 6.20e-21 &  5.95e-19  &  9.50e-20 \\
5.1328 & P 43 & 8005 & 5.94e-20 & 1.36e-20 &  9.06e-19  &  2.07e-19 \\
\enddata
\end{deluxetable*}
\begin{deluxetable*}{cccccccc}[h]
\tablecaption{ H$_2$O Line Fluxes \label{table:h2o_fluxes}}
\tablehead{
\colhead{$\lambda_0$} & \colhead{J K$_a$ K$_c$ } & \colhead{J K$_a$ K$_c$} &\colhead{ E$_\mathrm{up}/k$} & \colhead{Observed Flux} & \colhead{Error} & \colhead{Intrinsic Flux} & \colhead{Error}\\
\colhead{[$\mu$m]} &(upper)&(lower)&[K] &[$\mathrm{W\,m^{-2}}$] &[$\mathrm{W\,m^{-2}}$] &\colhead{[$\mathrm{W\,m^{-2}}$]}&\colhead{[$\mathrm{W\,m^{-2}}$]} }
\startdata
5.8227 & 6  1  6 & 5  0  5 & 2939 & 1.38e-20 & 1.66e-21 &  1.13e-18 &  1.37e-19 \\
5.8304 & 6  0  6 & 5  1  5 & 2938 & 5.53e-21 & 1.20e-21 &  4.73e-19  &   1.03e-19 \\
5.8605 & 2  2  0 & 1  1  1 & 2508 & 1.55e-20 & 3.55e-21 &   1.40e-18  &   3.20e-19 \\
5.9353 & 4  1  4 & 3  0  3 & 2621 & 1.95e-20 & 1.24e-21 &   2.11e-18  &   1.34e-19 \\
5.9902 & 3  1  3 & 2  0  2 & 2503 & 2.66e-20 & 2.98e-21 &   2.89e-18  &   3.24e-19 \\
6.0139 & 2  2  1 & 2  1  2 & 2507 & 1.02e-20 & 2.30e-21 &   1.05e-18  &   2.38e-19 \\
6.1432 & 2  0  2 & 1  1  1 & 2396 & 2.62e-20 & 6.80e-21 &   1.98e-18  &   5.16e-19 \\
6.1593 & 2  1  1 & 2  0  2 & 2437 & 8.58e-21 & 8.92e-22 &   6.23e-19  &   6.48e-20 \\
6.1854 & 1  1  0 & 1  0  1 & 2360 & 5.20e-20 & 1.93e-21 &   3.50e-18  &   1.30e-19 \\
6.3444 & 1  0  1 & 1  1  0 & 2329 & 1.08e-19 & 3.08e-21 &   4.95e-18  &   1.41e-19 \\
6.3703 & 2  0  2 & 2  1  1 & 2396 & 4.27e-20 & 4.50e-21 &   1.79e-18  &   1.88e-19 \\
6.3903 & 1  1  1 & 2  0  2 & 2352 & 2.25e-20 & 1.98e-21 &   8.69e-19  &   7.64e-20 \\
6.4163 & 3  0  3 & 3  1  2 & 2492 & 8.59e-20 & 2.14e-20 &   3.03e-18  &   7.55e-19 \\
6.4335 & 5  1  4 & 5  2  3 & 2879 & 2.29e-20 & 7.24e-21 &   7.77e-19  &   2.46e-19 \\
6.5552 & 4  2  3 & 4  3  2 & 2745 & 6.59e-20 & 1.04e-20 &   1.84e-18  &   2.92e-19 \\
6.5901 & 3  1  3 & 4  0  4 & 2503 & 3.07e-20 & 4.41e-21 &   9.31e-19  &   1.33e-19 \\
6.6124 & 6  2  5 & 6  3  4 & 3110 & 1.66e-20 & 3.31e-21 &   5.67e-19  &   1.13e-19 \\
6.6720 & 1  1  1 & 2  2  0 & 2352 & 5.53e-20 & 4.27e-21 &   2.72e-18  &   2.10e-19 \\
6.6834 & 4  1  4 & 5  0  5 & 2621 & 5.41e-20 & 4.31e-21 &   2.74e-18  &   2.18e-19 \\
6.7234 & 2  1  1 & 3  2  2 & 2437 & 2.26e-20 & 4.46e-21 &   1.23e-18  &   2.43e-19 \\
6.7745 & 5  1  5 & 6  0  6 & 2767 & 2.89e-20 & 8.37e-21 &   1.82e-18  &   5.28e-19 \\
6.7865 & 5  0  5 & 6  1  6 & 2764 & 4.87e-20 & 1.39e-20 &   3.13e-18  &   8.91e-19 \\
6.7932 & 3  1  2 & 4  2  3 & 2550 & 4.72e-20 & 8.90e-21 &   3.02e-18  &   5.70e-19 \\
6.8264 & 2  1  2 & 3  2  1 & 2413 & 6.89e-20 & 4.56e-21 &   4.53e-18 &   3.00e-19 \\
6.8528 & 4  1  3 & 5  2  4 & 2698 & 2.66e-20 & 7.77e-21 &   1.64e-18  &   4.78e-19 \\
6.9933 & 3  2  2 & 4  3  1 & 2610 & 3.24e-20 & 8.22e-21 &   1.19e-18  &   3.02e-19 \\
7.0239 & 3  1  3 & 4  2  2 & 2503 & 5.19e-20 & 8.89e-21 &   1.73e-18  &   2.96e-19 \\
7.1469 & 4  2  3 & 5  3  2 & 2745 & 7.04e-20 & 1.21e-20 &   1.50e-18  &   2.58e-19 \\
7.1712 & 4  3  2 & 5  4  1 & 2884 & 4.77e-20 & 1.14e-20 &   9.86e-19  &   2.36e-19 \\
7.2071 & 4  4  1 & 5  5  0 & 3064 & 6.88e-20 & 1.36e-20 &   1.41e-18  &   2.80e-19 \\
7.2792 & 5  3  2 & 6  4  3 & 3065 & 5.45e-20 & 9.48e-21 &   9.41e-19  &   1.64e-19\\
\enddata
\end{deluxetable*}

\end{document}